\documentclass[12pt]{article}
\usepackage{fullpage}
\usepackage{epsfig}

\def\be{\begin{equation}}
\def\ee{\end{equation}}
\def\beq{\begin{eqnarray}}
\def\eeq{\end{eqnarray}}
\def\msl {m_{slep}}

\def\gsim{\:\raisebox{-1.1ex}{$\stackrel{\textstyle>}{\sim}$}\:}

\def\E{${\rm E}\!\!\!\!/$}
\def\Et{${{\rm E}\!\!\!\!/}_{\rm T}$}

\begin{document}

\begin{center}
{\Large\bf Event-shape of dileptons plus missing energy at a linear
collider as a SUSY/ADD discriminant} \\[1cm]
{\large Probir Roy} \\[.5cm]
Tata Institute of Fundamental Research, Mumbai, India \\[1cm]
%(Talk given at the LCWS06 Workshop, Bangalore, March 2006)
\end{center}
\bigskip

\begin{enumerate}
\item[{}] {\bf $\bullet$ Objective $\bullet$ Signal cross section $\bullet$
SM background and chosen cuts \\ $\bullet$ Event-shape variables $\bullet$
Results and discussion}
\end{enumerate}
\bigskip

\noindent {\bf $\bullet$ Objective} -- This talk is based on work done
with Partha Konar [1].  New physics is widely expected to emerge at
TeV energies on the basis of naturalness, gauge hierarchy and WIMP
dark matter considerations.  Among possible scenarios, supersymmetry
(SUSY) [2] and large extra dimensions [3] of the
Arkani--Hamed--Dimopoulos--Dvali (ADD) model [4] are in the limelight.
This is because they promise a large number of new states to be
explored at the LHC as well as ILC and CLIC.  The signature of these
states is the occurrence of multilepton, multijet events with a large
missing energy \E \ or missing transverse energy \Et.  The question of
discrimination between the two scenarios on the basis of such events
is thus an important issue.
\bigskip

\noindent {\bf $\bullet$ Signal cross section} -- We consider the
lepton sector where the LHC will not be a powerful probe.  Hence we
zero in on ILC ($\sqrt{s} = 500$ GeV) and CLIC ($\sqrt{s} = 3$ TeV).
Our process is $e^+e^- \rightarrow \ell^+ \ell^-$\E \ where $\ell$ is
an electron or a muon.  For SUSY, we take the MSSM with the parameters
$\tan\beta$, $m_{\tilde e_{L,R}}$, $m_{\tilde\mu_{L,R}}$, $\mu$, $M_1$
and $M_2$ in standard notation, with all mass parameters expected to
be $\gsim {\cal O}$ (100 GeV).  The production of a pair of charged sleptons
$\tilde \ell^\pm_{L,R}$ and their subsequent decays into
$\ell^\pm\tilde\chi^0_1$ lead to our signal in this scenario.  Signal
sensitivity to $\tan\beta$ turns out to be very mild and {\it we fix
the latter at 10}.  Thus we have a 4-parameter SUSY model.  Turning to
the ADD case, we take $d$ extra spatial dimensions (with $d =
2,3,4,5$), all compactified on a $d$-torus with the same radius $R_c$
of compactification for each.  All Standard Model fields are assumed
to lie on a 3-brane while only gravity is taken to propagate in the
bulk.  The parameters $d,R_c$ and the fundamental `string' scale $M_S$
in higher dimensions are related by $M^{2+d}_S = (4\pi)^{d/2}
\Gamma(d/2)G^{-1}_N R^{-d}_c$, $G_N$ being Newton's constant, so
that one can take $d$ and $M_S$ to be the two independent parameters
of this model.

A reliable discriminant between the SUSY and ADD scenarios, apart from
being a measurable quantity, needs to have robust features
distinguishing between them.  Such is not the case with the lepton
energy spectrum here.  For appropriate parameters and with ISR
corrections, the famous box-shaped lepton energy spectrum from slepton
decay can get squeezed [5] into a hump (Fig. 1a), not too unlike that
in the ADD case (Fig. 1b).  Lepton angular 

%\vspace*{6cm}
\begin{center}
\epsfig{file=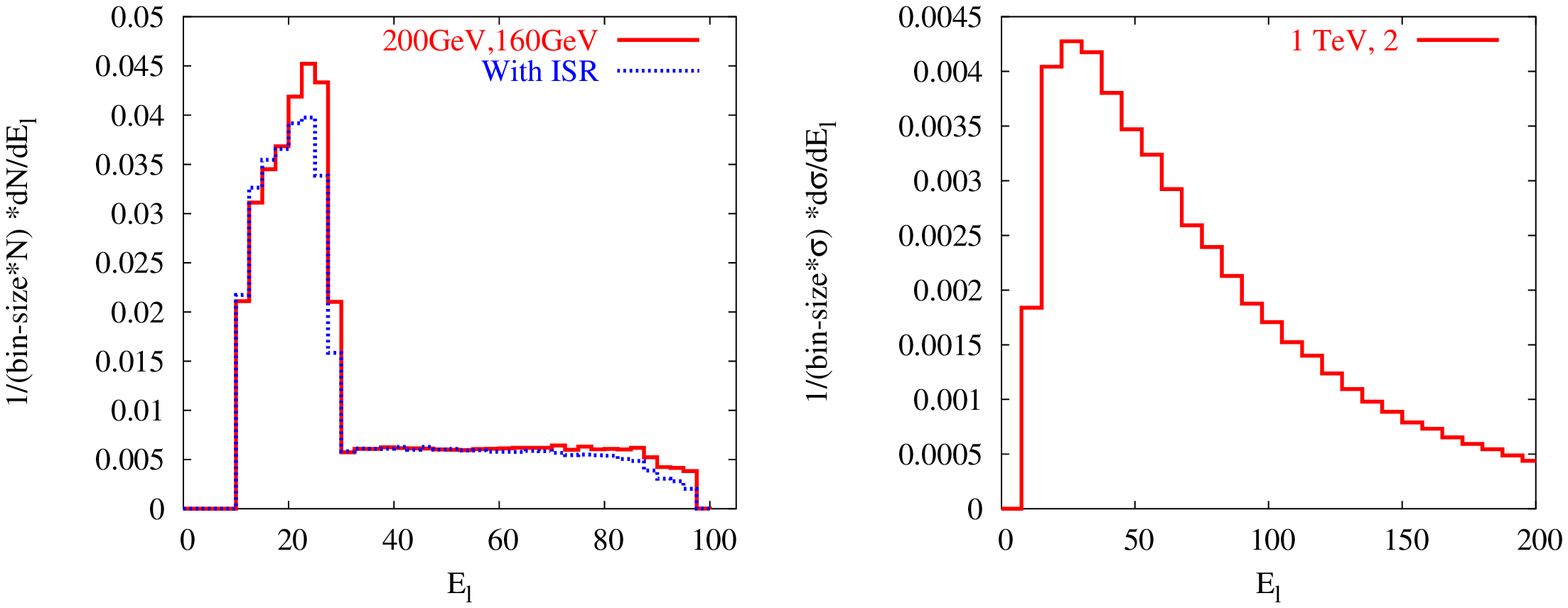,height=6cm}
\end{center}
\begin{center}
Fig. 1. Lepton energy spectrum : SUSY (ADD) in left (right) panel
with $M_2$ and $M_1$ \\ ($M_S$ and $d$) specified.
\end{center}

\noindent distributions also tend to be flat and somewhat
similar in both scenarios for most of the allowed range. In contrast,
event-shape distributions like those of sphericity and thrust are
robust with respect to ISR/FSR corrections and differ significantly
for the two scenarios.

All lowest order diagrams, relevant to the process $e^+e^- \rightarrow
\ell^+ \ell^-$ \E \ in the ADD (SUSY) case are shown in the left
(right) panel of Fig. 2.  For the former, one can write 
\medskip
 
\hspace*{5.5cm} ADD \hspace*{2cm} SUSY
\begin{center}
\epsfig{file=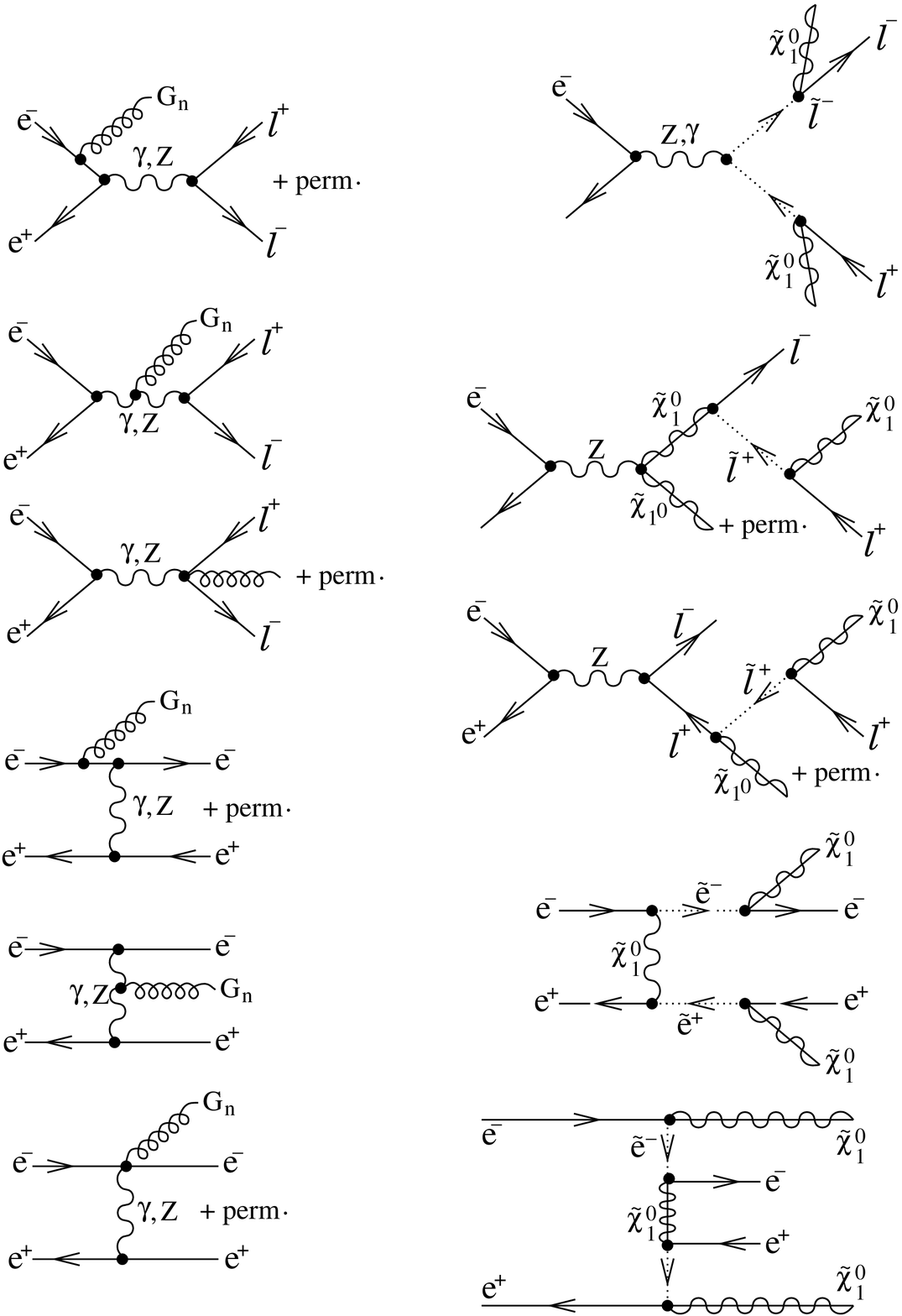,height=7cm}
\end{center}

\begin{center}
Fig. 2. Lowest order diagrams for our process in the two cases.
\end{center}
\[
\sigma(e^+e^- \rightarrow \ell^+\ell^- {E\!\!\!\!/}) = \Sigma_n
\sigma(e+e^- \rightarrow \ell^+ \ell^- G_n) \simeq \int^{\sqrt{s}}_0
dm \ \sigma(m) \left[2 R^d_c m^{d-1} (4\pi)^{-d/2}/\Gamma(d/2)\right],
\]
$m$ being the mass of the graviton mode with $\sigma(m)$ being the
corresponding production cross section.  The latter has been
calculated with the subroutine HELAS.  For the SUSY case, the rate for
the process $e^+e^- \rightarrow \tilde\ell^+_{L,R} \tilde\ell^-_{L,R}
\rightarrow \ell^+ \ell^- \tilde\chi^0_1 \tilde\chi^0_1$ has been
calculated by using the package COMPHEP.  Typical cross sections for
the two scenarios, computed with cuts described in the next section,
are listed in Table 1.  Values for SUSY and ADD show considerable
overlap both for ILC (upper half) and CLIC (lower half).

%\begin{table}
$$
\begin{array}{|c|c|c|c|c|c|c|c|c|c|c|c|}
\hline
\multicolumn{6}{|c|}{\bf \sigma_{SUSY} [fb]}  &|&
\multicolumn{5}{|c|}{\bf \sigma_{ADD} [fb]}\\
\cline{1-6} \cline{8-12}
\multicolumn{2}{|c|}{\bf \tan \beta = 10}    &
\multicolumn{4}{|c|}{\bf \msl [GeV]}  &|&   \multicolumn{4}{|c|}{\bf
M_S [TeV]}  & \\
\cline{1-6} \cline{8-11}
{\bf M_2, M_1 [GeV]} & {\bf \mu [GeV]} & \bf 155  & \bf 205  & \bf 225
 & \bf 245     &|&    \bf .75  & \bf 1.0    & \bf 1.5  & \bf 2.0  &
\bf d   \\ 
\cline{1-6} \cline{8-12}                
\bf 200, 100   &  \bf -400    &  427   &  164     &   59     &   7.8
&|&   1090  &  345      &  68    &   22 & \bf 2   \\
\cline{1-6} \cline{8-12}                 
\bf 300, 150   &  \bf -400    &  144   &  137     &   75     &   19
&|&   455   &  108      &  14    &   3.3 & \bf 3    \\
\cline{1-6} \cline{8-12}        
\bf 400, 200   &  \bf -150    &   92   &  40      &   13     &   0.6
&|&  202   &  36       &  3.2   &   0.6 & \bf 4  \\
\cline{1-6} \cline{8-12}
\bf 400, 200   &  \bf -100    &   79   &  32      &   6.9    &   0.3
&|&   97    &  13       &  0.8   &  0.1 & \bf 5\\
\hline
\end{array}
$$
$$
\begin{array}{|c|c|c|c|c|c|c|c|c|c|c|c|}
\hline
\multicolumn{6}{|c|}{\bf \sigma_{SUSY} [fb]}  &|&
\multicolumn{5}{|c|}{\bf \sigma_{ADD} [fb]}\\
\cline{1-6} \cline{8-12}
\multicolumn{2}{|c|}{\bf \tan \beta = 10}    &
\multicolumn{4}{|c|}{\bf \msl [GeV]}  &|&   \multicolumn{4}{|c|}{\bf
M_S [TeV]}  &
\\
\cline{1-6} \cline{8-11}
{\bf M_2, M_1 [GeV]} & {\bf \mu [GeV]} & \bf 700  & \bf 800  & \bf 900
 & \bf 1000     &|&    \bf 4.5  & \bf 5.0    & \bf 5.5  & \bf 6.0  &
\bf d   \\
\cline{1-6} \cline{8-12}
\bf 200, 100   &  \bf -500    &   24   &   19     &   15     &    11
&|&   124  &  81      &  56    &   39 & \bf 2   \\
\cline{1-6} \cline{8-12}
\bf 400, 190   &  \bf -500    &   22   &   18     &   15     &   11
&|&   58   &  34      &  21    &   14 & \bf 3    \\
\cline{1-6} \cline{8-12}
\bf 600, 290   &  \bf -500    &   21   &   16     &   13     &   10
&|&   31   &  16       &  9.2   &   5.5 & \bf 4  \\
\cline{1-6} \cline{8-12}
\bf 800, 380   &  \bf -500    &   21   &   18     &   12     &    8
&|&   17    &  8.3       &  4.2   &  2.3 & \bf 5\\
\hline
\end{array}
$$
\begin{center}
SM bkgd $\sim$ 36 fb (72 fb)
\end{center}
\begin{center}
Tabl 1. Cross sections for various parameters of the two scenarios
\end{center}
\bigskip

\noindent {\bf $\bullet$ SM background and chosen cuts} -- The main
background to our signal comes from the reactions $e^+e^- \rightarrow
\ell^+ \ell^- Z$, $Z \rightarrow \nu_\ell \bar\nu_\ell$ and $e^+e^-
\rightarrow W^+ W^-$, $W \rightarrow \ell \nu_\ell$.  The first can be
eliminated by a missing mass cut clearly excluding $M_Z$.  The second
is kinematically reconstructible modulo a 2-fold ambiguity and can
then be explicitly subtracted.  The signal to background ratio gets
further enhanced on account of our cuts chosen as follows.  (1) Each
$\ell$ must be at least $10^\circ$ from the beam pipe to control
beamsstrahlung effects and collinear singularities from $t$-channel
photon exchange.  (2) For each $\ell$, $p^\ell_T$ must exceed 10 GeV
(ILC) or 20 GeV (CLIC).  (3) The corresponding acceptance lower limits
for $p_T^{miss}$ are chosen as 15 GeV and 25 GeV respectively.  (4)
The isolation criterion $\Delta R \equiv (\Delta \eta^2 + \Delta
\phi^2)^{1/2} > 0.2$ is chosen.  (5) The opening angle acceptance
range is taken as $5^\circ < \theta_{\ell^+\ell^-} < 175^\circ$.  
(6) The missing mass cut is chosen to be $M_{miss} > 150$ GeV (ILC),
450 GeV (CLIC).  With these cuts, the SM background is about 36 fb
(ILC), 72 fb (CLIC) to be compared with the signal numbers in Table
1.  For an integrated luminosity of 100 fb$^{-1}$ (ILC), 1000
fb$^{-1}$ (CLIC), a minimum signal cross section of 1.8 fb, 0.8 fb
would achieve $S/\sqrt{B} \simeq 3$.
\bigskip

\noindent {\bf $\bullet$ Event-shape variables} -- The idea of using
event-shape variables arises from the following expectation.  Decay
products from a slepton pair, produced not far from threshold, are
likely to be more isotropic as compared with the somewhat more spiked
configurations of bremsstrahlunglike graviton emission in the ADD
case.  We define a sphericity tensor $S_{ij}$ and a scalar parameter
thrust $T$ as 
\[
S_{ij} = {p^i_{\ell^+} p^j_{\ell^+} + p^i_{\ell^-} p^j_{\ell^-} \over
{\bf p}^2_{\ell^+} + {\bf p}^2_{\ell^-}}, \ T = {\rm max} {\hat{\bf n}
\cdot ({\bf p}_{\ell^+} + {\bf p}_{\ell^-}) \over |{\bf p}_{\ell^+}|
+ |{\bf p}_{\ell^-}|}, 
\]
where the thrust axis until vector $\hat{\bf n}$ is chosen to maximize
the numerator of $T$.  The allowed range for the latter is $1/2 \leq T
\leq 1$ and a spiked (isotropic) event has $T \sim 1 \ (1/2)$.  On the
other hand, if $\lambda_{1,2,3}$ are the eigenvalues of $S_{ij}$,
defined with $\lambda_1 \geq \lambda_2 \geq \lambda_3 \geq 0$,
$\lambda_1 + \lambda_2 + \lambda_3 = 1$, the sphericity $S$ of the
event can be defined as
\[
S = {3\over2} (\lambda_2 + \lambda_3)
\]
with $0 \leq S \leq 1$, where $S = 1,0$ for an ideally spherical,
linear event.  For our process, the planar nature of two body
production implies that $\lambda_3 = 0$.  Thus the shape of an
isotropic event is circular (rather than spherical) for $S_{\rm max} =
3/4$.  However, ISR/FSR effects can, in principle, push $S$ beyond the
maximum and towards unity.
\bigskip

\noindent {\bf $\bullet$ Results and discussion} -- The resultant
sphericity and thrust distributions for the two scenarios, as relevant
to ILC (CLIC) is shown in Fig. 3 (Fig. 4).  We have cross-checked 

\vspace*{4cm}

\begin{figure}[h]
\includegraphics{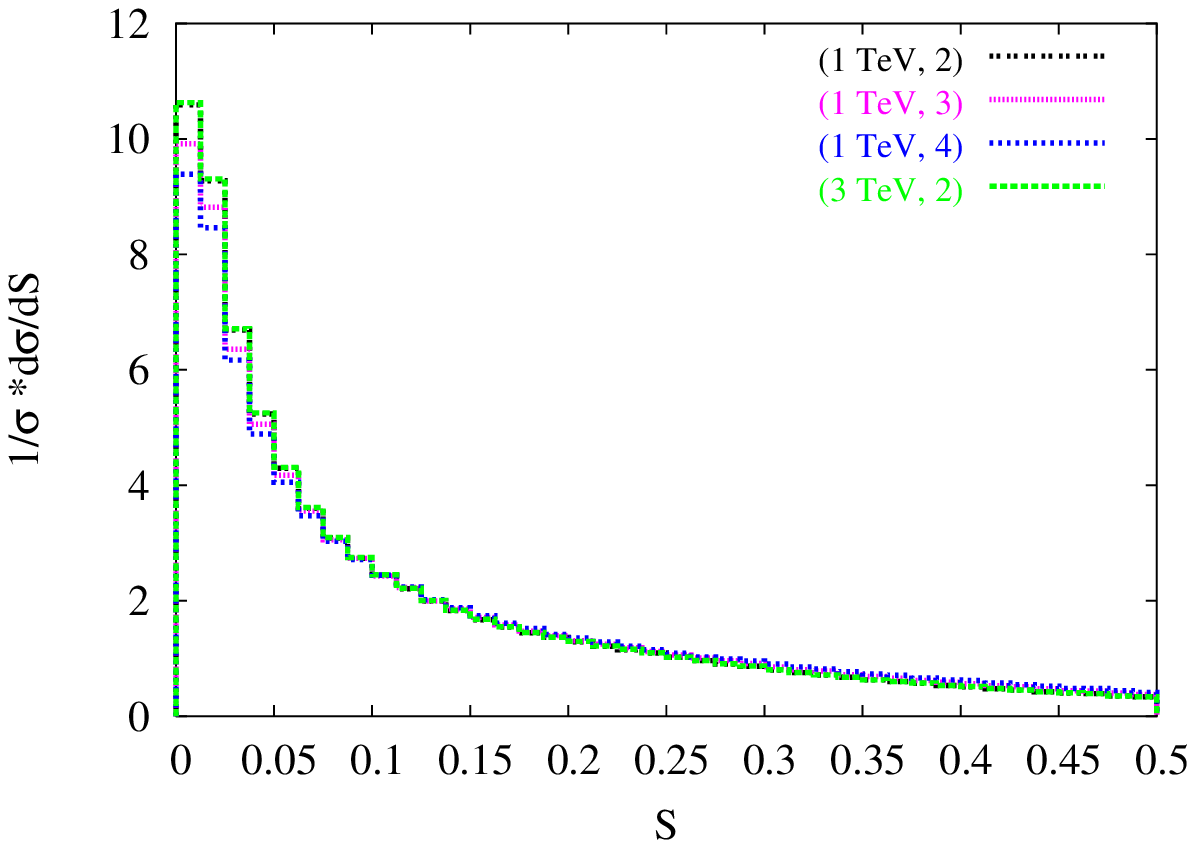} 
\label{fig:Fig1.eps}
\includegraphics{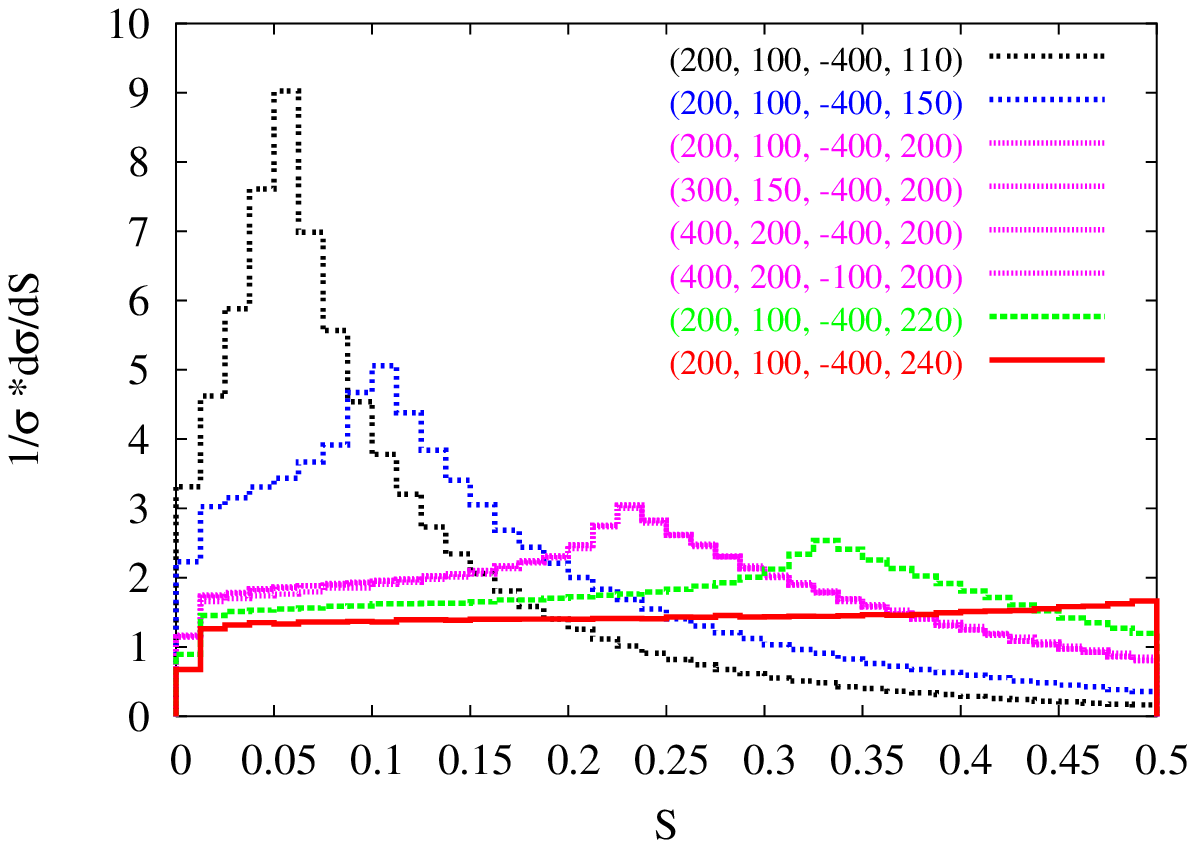} 
\label{fig:Fig2.eps}
\end{figure}

\vspace*{3.5cm}

\begin{figure}[h]
\includegraphics{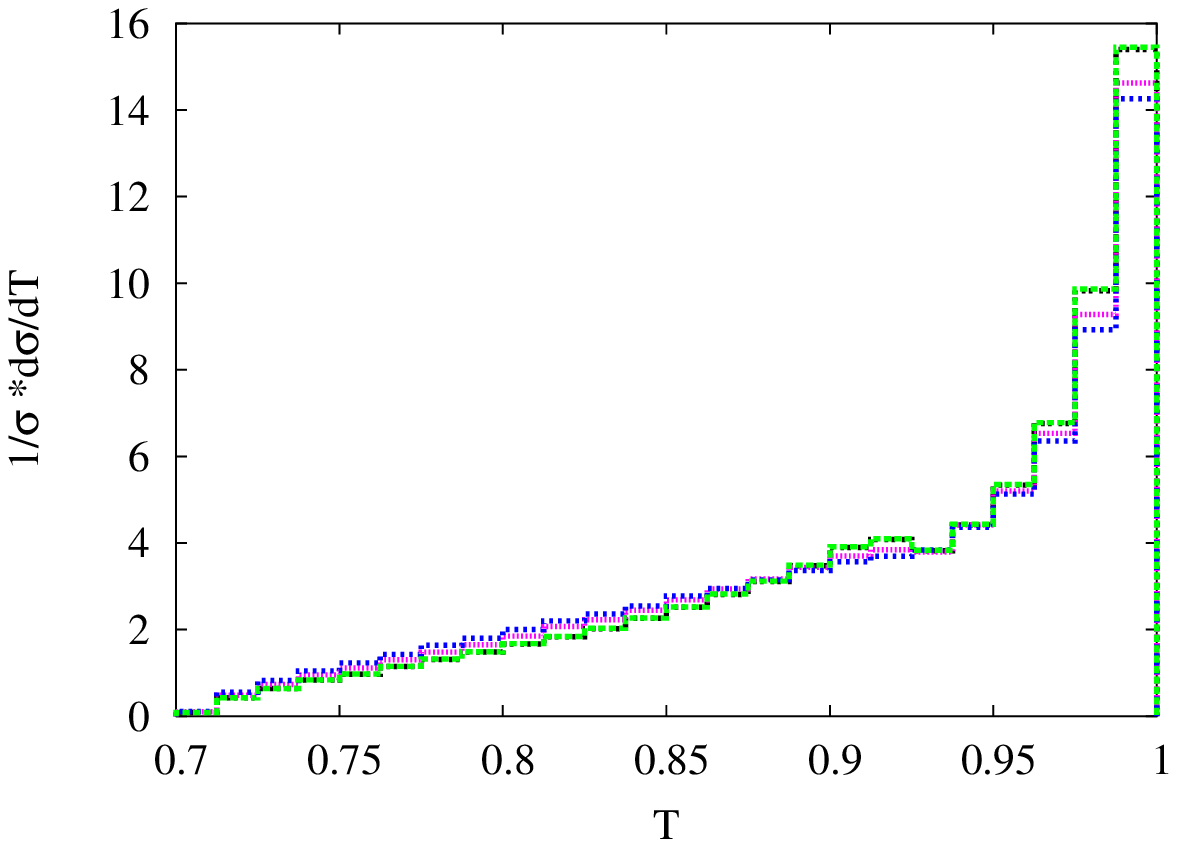} 
\label{fig:Fig3.eps}
\includegraphics{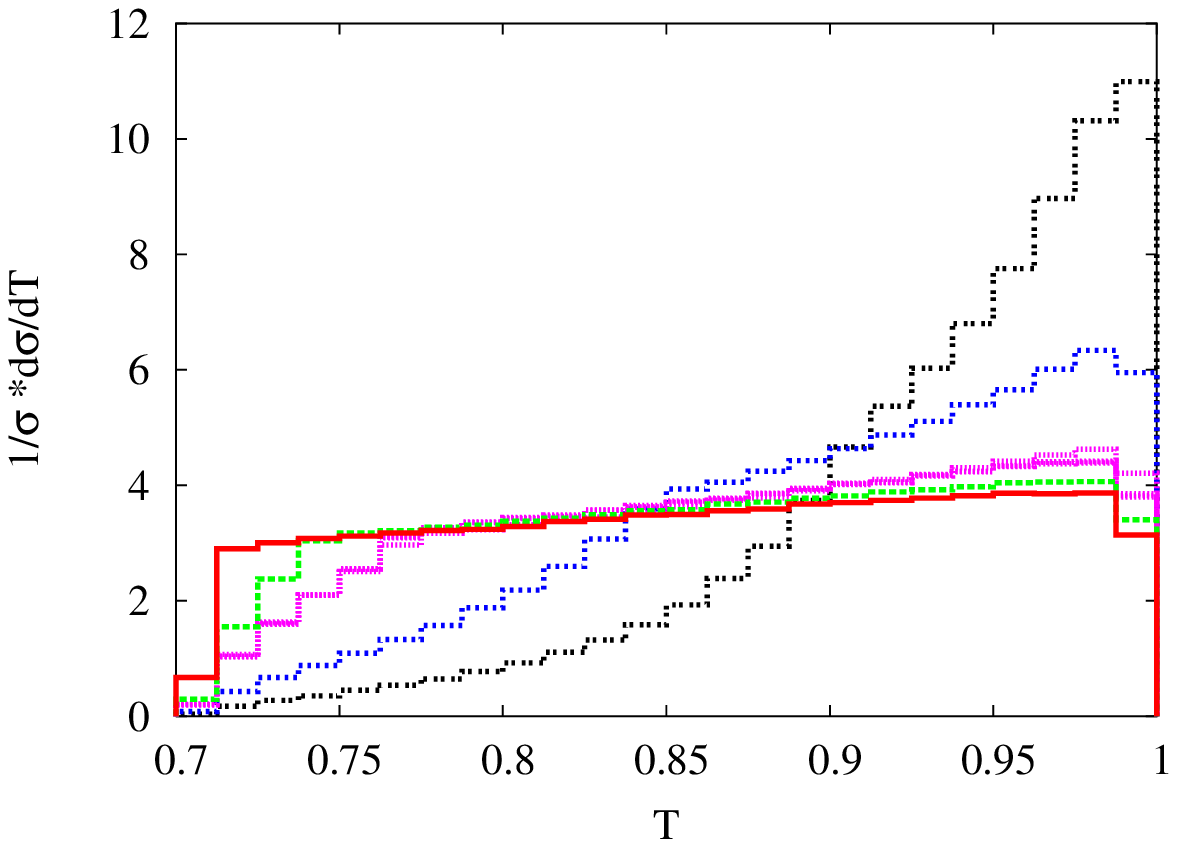} 
\label{fig:Fig4.eps}
\end{figure}

\begin{center}
Fig. 3. Sphericity (upper panels) and thrust (lower panels)
distributions \\ for ADD (left) and SUSY (right) at ILC.
\end{center}

\vspace*{6cm}

\begin{figure}[h]
\includegraphics{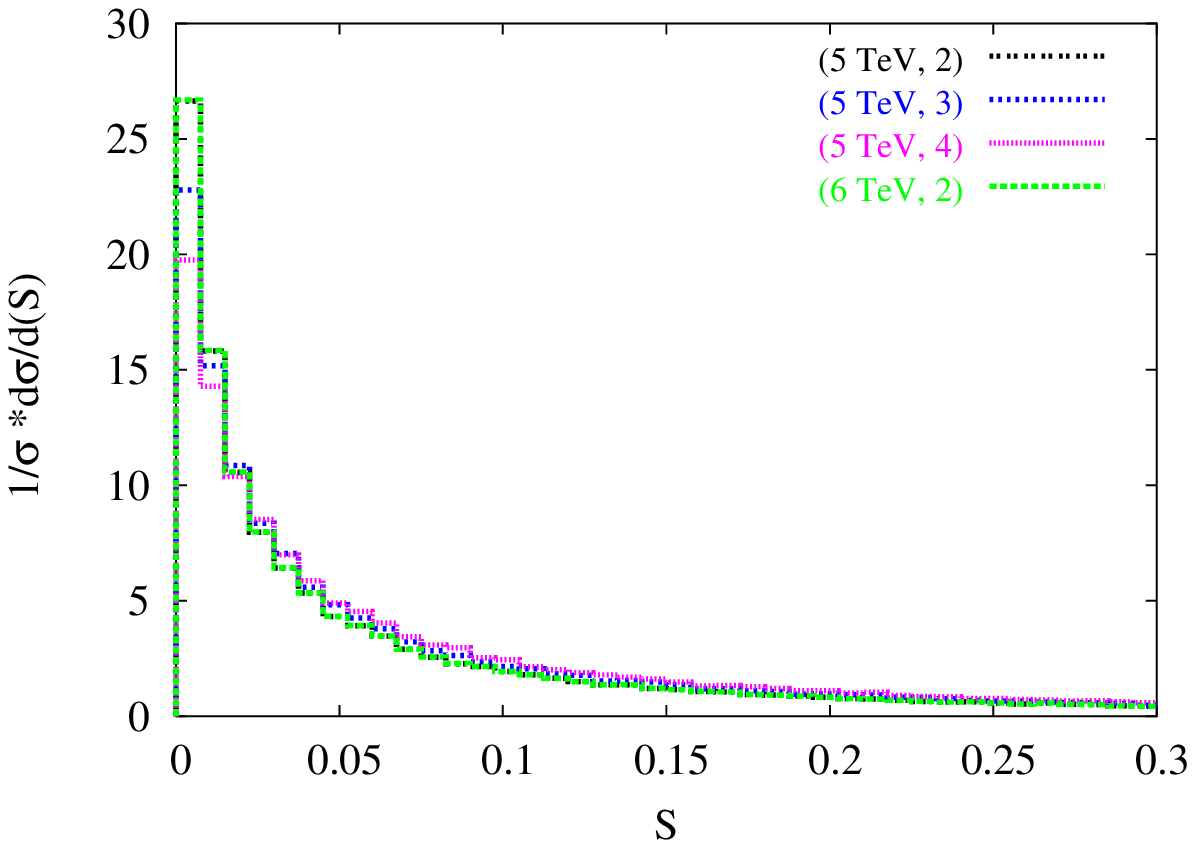} 
\label{fig:Fig5.eps}
\includegraphics{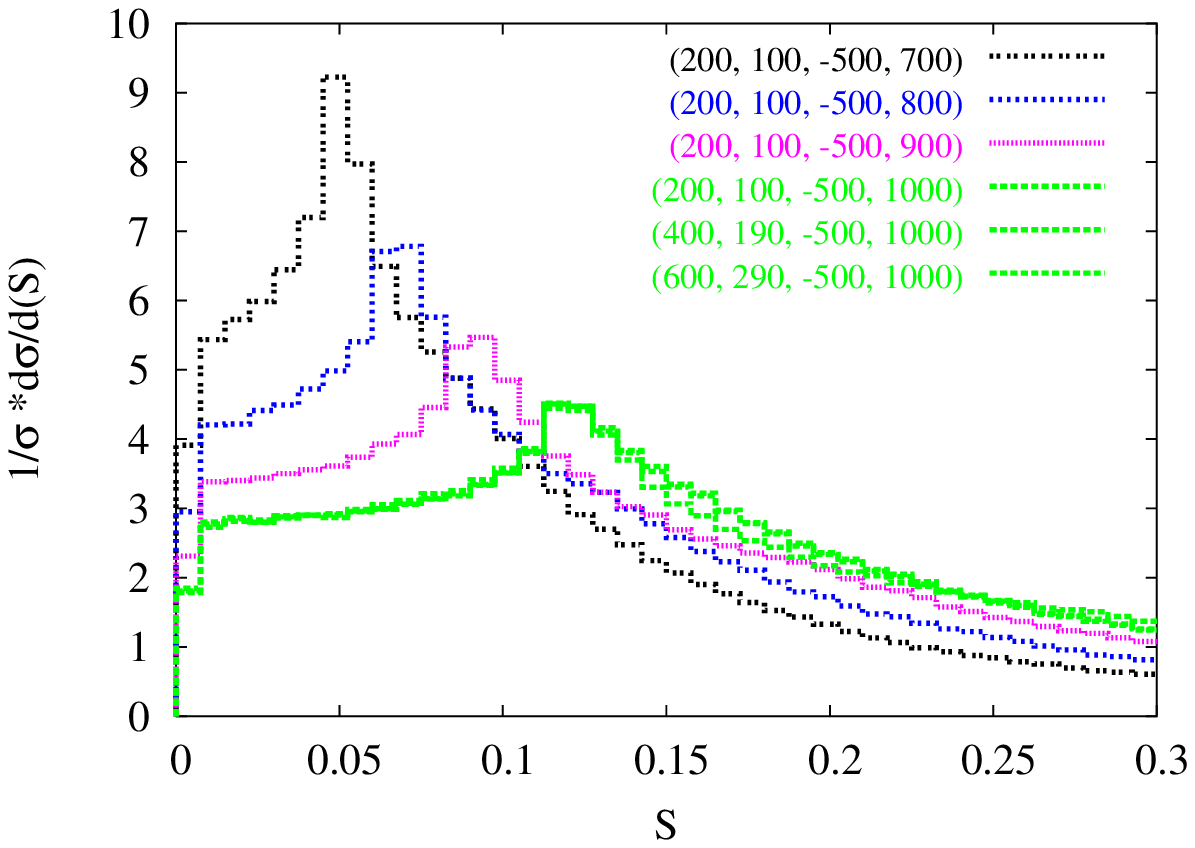} 
\label{fig:Fig6.eps}
\end{figure}

\vspace*{4.5cm}

\begin{figure}[h]
\includegraphics{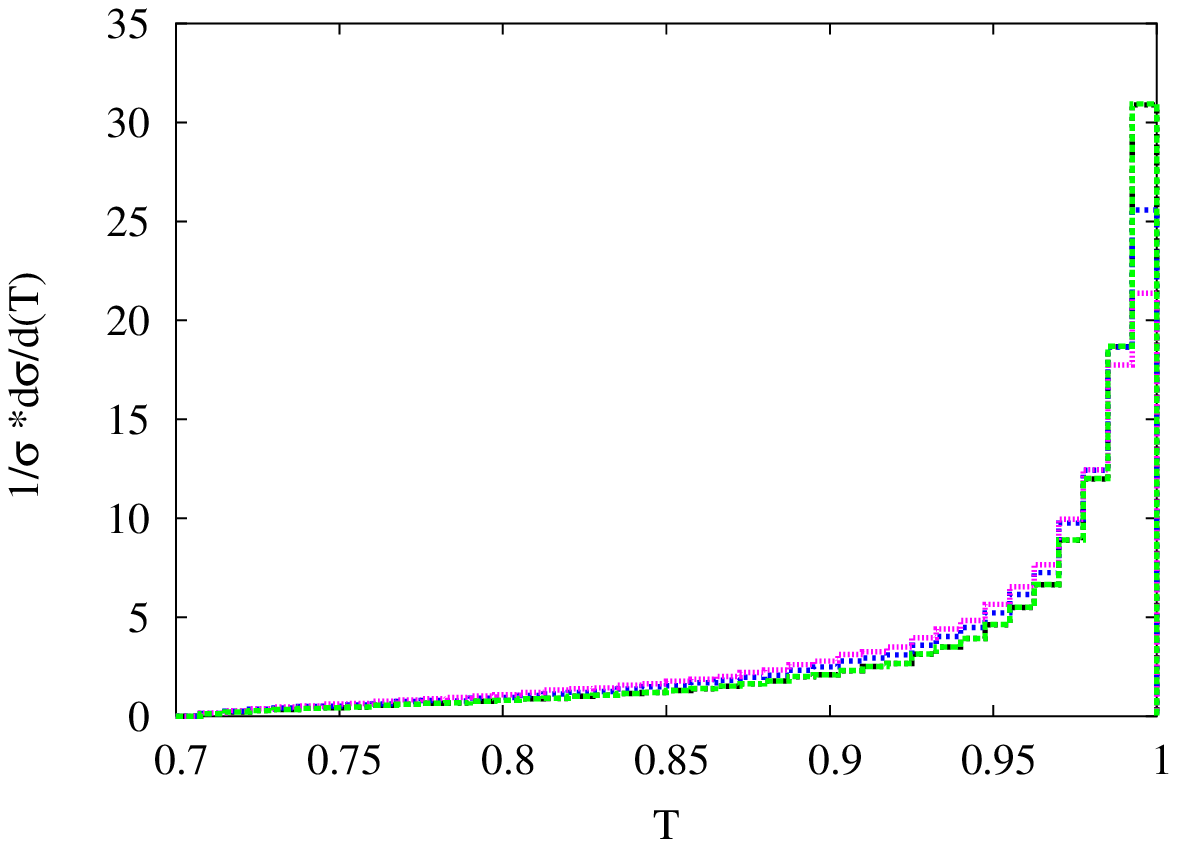} 
\label{fig:Fig7.eps}
\includegraphics{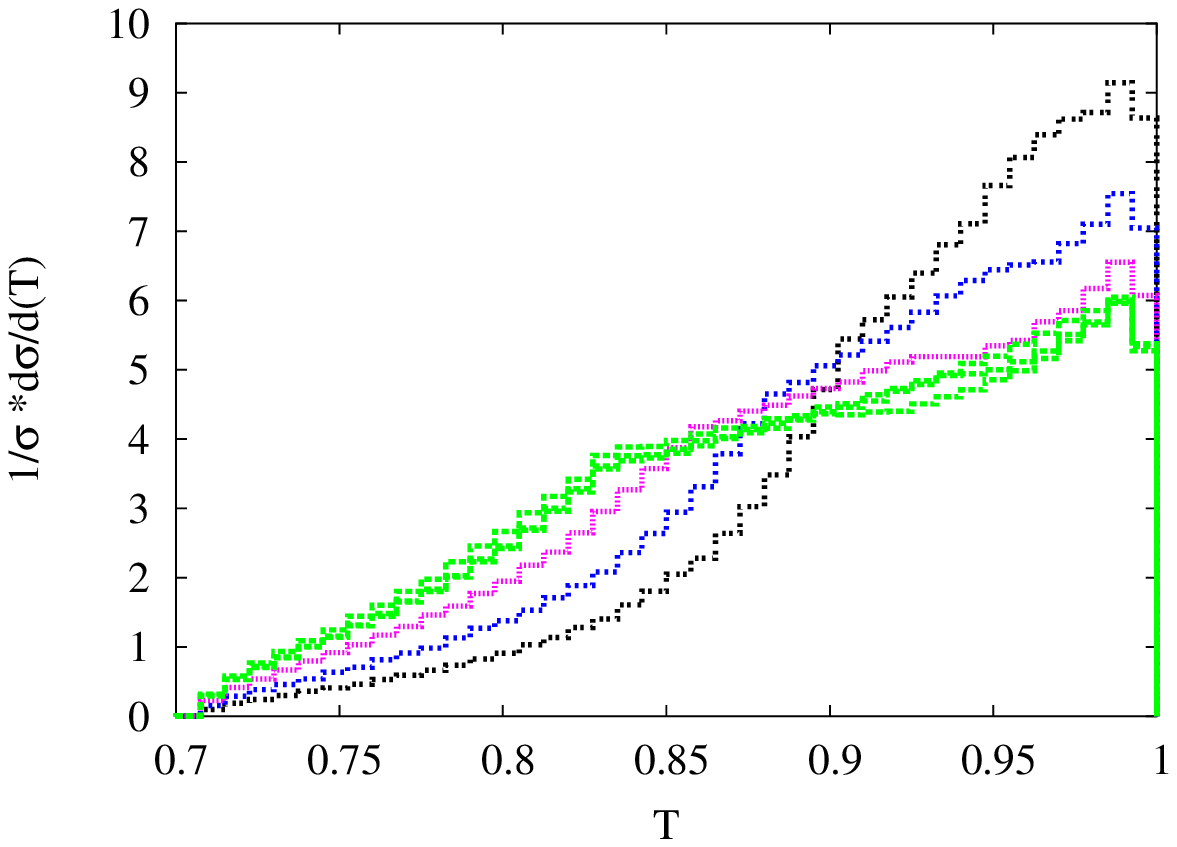} 
\label{fig:Fig8.eps}
\end{figure}

\begin{center}
Fig. 4. Sphericity (upper panels) and thrust (lower panels)
distributions \\ for ADD (left) and SUSY (right) at CLIC.
\end{center}

\noindent the SUSY plots by redoing [1] the calculation
in PYTHIA with ISR/FSR effects taken into account.  The observed
changes are small, showing the robustnes of these event-shape
variables with respect to such corrections.

An examination of these plots clarifies the distinction between the
two scenarios. Both $S$ and $T$ distributions are flatter in the SUSY
case, showing structure in terms of a peak in $S$ and a break in $T$.
In contrast, they are monotonic for ADD with maxima at $S=0$ and $T=1$
(spiked event), followed by a continuous fall and rise respectively.
It is a fact that the discrimination is more spectacular via $S$ than
via $T$.  In the SUSY case, the location of the sphericity peak is uniquely
correlated with the slepton mass $m_{\tilde\ell}$, being insensitive
to other MSSM parameters.  This is demonstrated in the scatter plot of
Fig. 5 displaying the cross section against the said location.  In
contrast, the maxima for all ADD parametric choices are strictly at
$S=0$. 

%\vspace*{6cm}
\begin{center}
\epsfig{file=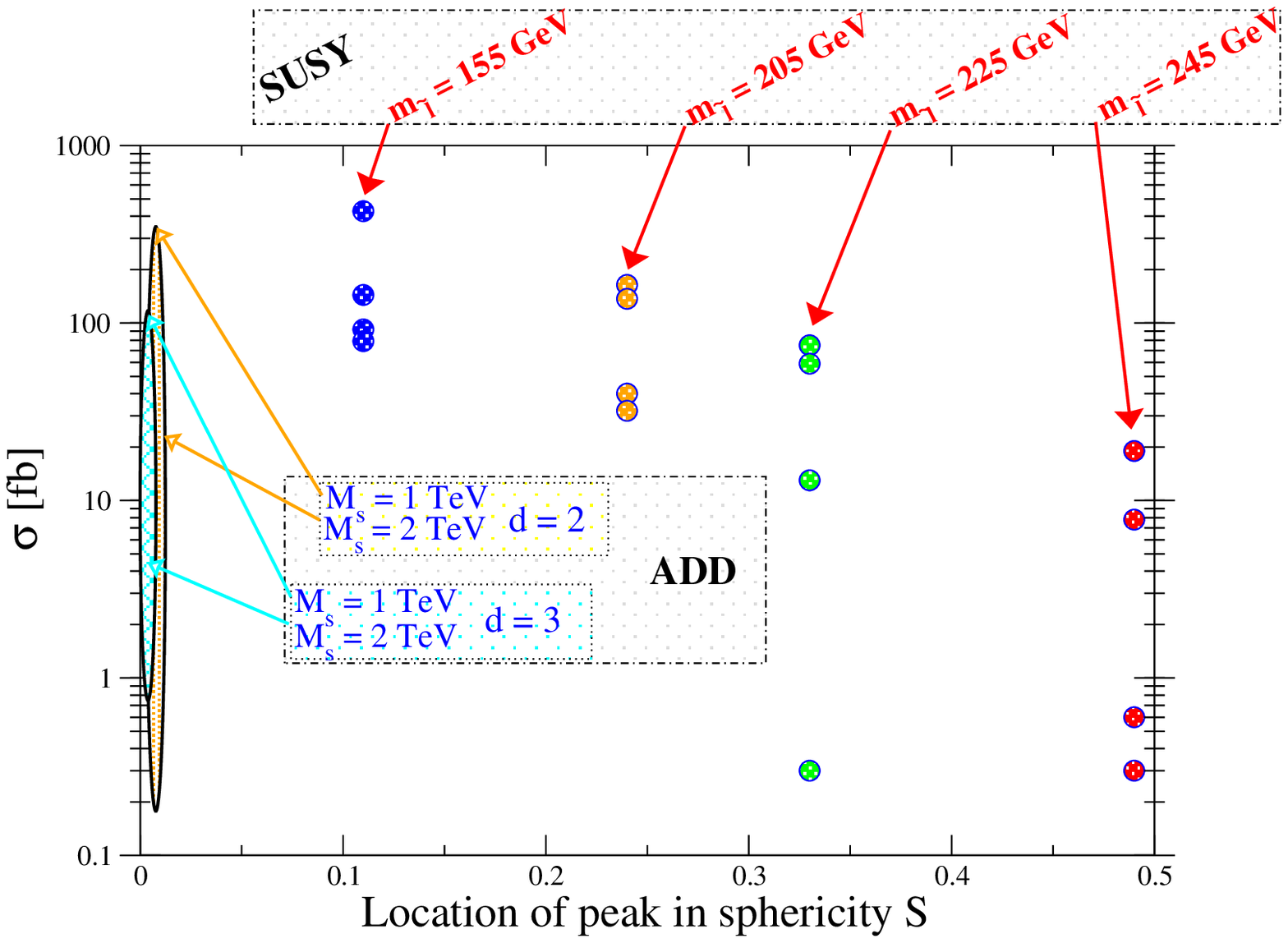,height=6cm}
\end{center}
\begin{center}
Fig. 5. Sphericity peak location
\end{center}

In sum, a clear discrimination between SUSY and ADD will be possible in
$e^+e^- \rightarrow \ell^+ \ell^-$ \E \ at a linear collider by means
of ISR/FSR-insensitive sphericity distributions.  A peaked structure,
with the peak location uniquely specifying the slepton mass,
characterizes SUSY.  In contrast, a structureless monotonic falloff
from a maximum at $S=0$ is the hallmark of ADD.

\begin{enumerate}
\item[{[1]}] P. Konar and P. Roy, Phys. Lett. {\bf B634} (2006) 295.
\item[{[2]}] M. Drees, R.M. Godbole and P. Roy, {\it Theory and
Phenomenology of Sparticles}, World Scientific, Singapore (2004).
\item[{[3]}] A. Perez-Lorenzana, J. Phys. Conf. Ser. {\bf 18} (2005)
224. 
\item[{[4]}] I. Antoniadis, Phys. Lett. {\bf B246} (1990) 377.
N. Arkani-Hamed {\it et. al.}, Phys. Lett. {\bf B429} (1998) 263;
Phys. Rev. {\bf D59} (1999) 086004.
\item[{[5]}] M. Battaglia {\it et. al.}, JHEP {\bf 0507} (2005) 033. 
\end{enumerate}

\end{document}